\documentclass[pdf,twocolumn,superscriptaddress,floatfix,showpacs,amsmath,longbibliography,nofootinbib,amssymb]{revtex4-1}
\usepackage[colorlinks=true,citecolor=blue,linkcolor=blue]{hyperref}
\usepackage[usenames]{color}
\usepackage{subfigure}
\usepackage{dcolumn}
\usepackage{mathrsfs}% Align table columns on decimal point
\usepackage{bm}% bold math
\usepackage{amsmath,amssymb,epsfig,float,graphics}
\newcommand{\bk}{\mathbf{k}}
\newcommand{\br}{\mathbf{r}}

\newcommand{\bx}{\mathbf{x}}

\begin{document}
\baselineskip=0.45 cm

\title{Direct Characteristic-Function Tomography of the Quantum States of Quantum Fields}

\author{Zehua Tian}
\email{tianzh@ustc.edu.cn}
\affiliation{CAS Key Laboratory of Microscale Magnetic Resonance and School of Physical Sciences, University of Science and Technology of China, Hefei 230026, China}
\affiliation{CAS Center for Excellence in Quantum Information and Quantum Physics, University of Science and Technology of China, Hefei 230026, China}

\author{Jiliang Jing}
%\email{jljing@hunnu.edu.cn}
\affiliation{Department of Physics, Key Laboratory of Low Dimensional Quantum Structures and Quantum Control of Ministry of Education, and Synergetic Innovation Center for Quantum Effects and Applications, Hunan Normal University, Changsha, Hunan 410081, P. R. China}

\author{Jiangfeng Du}
\affiliation{CAS Key Laboratory of Microscale Magnetic Resonance and School of Physical Sciences, University of Science and Technology of China, Hefei 230026, China}
\affiliation{CAS Center for Excellence in Quantum Information and Quantum Physics, University of Science and Technology of China, Hefei 230026, China}
\affiliation{Hefei National Laboratory, University of Science and Technology of China, Hefei 230088, China} 
\affiliation{School of Physics, Zhejiang University, Hangzhou 310027, China}

\begin{abstract}

Herein, we propose a novel strategy for implementing a direct readout of the symmetric characteristic function of the quantum states of quantum fields without the involvement of idealized measurements, an aspect that has always been deemed ill-defined in quantum field theory. This proposed scheme relies on the quantum control and measurements of an auxiliary qubit locally coupled to the quantum fields. By mapping the expectation values of both the real and imaginary parts of the field displacement operator to the qubit states, the qubit’s readout provides complete information regarding the symmetric characteristic function. We characterize our technique by applying it to the Kubo--Martin--Schwinger (thermal) and squeezed states of a quantum scalar field. In addition, we have discussed general applications of this approach to analogue-gravity systems, such as Bose--Einstein condensates, within the scope of state-of-the-art experimental capabilities. This proposed strategy may serve as an essential in understanding and optimizing the control of quantum fields for relativistic quantum information applications, particularly in exploring the interplay between gravity and quantum, for example, the relation to locality, causality, and information.
\\           \par \
\end{abstract}
\maketitle
\baselineskip=0.45 cm

\newpage
%%%%%%%%%%%%%%%%%%%%%%%%%%%%%%%%%%%%%%%%%%%%%%%%%%%%%%%%%%%%
\section{Introduction}
In quantum information theory (QIT), full characterization of unknown quantum states enables the diagnosis and enhancement of quantum control, potentially ushering a revolution in current technologies ranging from fundamental tests of physics \cite{RevModPhys.85.471, RevModPhys.86.1391, RevModPhys.90.025008, RevModPhys.89.035002, GW} to different quantum-information-processing tasks \cite{nielsen2010quantum, RevModPhys.84.621, RevModPhys.86.153, RevModPhys.74.145}. In recent years, research on quantum-state tomography (QST) in both discrete- \cite{2004LNP649P, PhysRevA.66.012303, PhysRevA.55.R1561, PhysRevLett.105.150401, PhysRevLett.74.4101, QT1, PhysRevLett.103.160503, QT2, QT3, PhysRevLett.122.110406, PhysRevLett.124.010405} and continuous-variable quantum systems \cite{2004LNP649P, PhysRevLett.122.110406, RevModPhys.81.299, PhysRevLett.120.090501, PhysRevLett.125.043602, PhysRevResearch.4.033220} has been conducted in both theoretical and experimental aspects using direct and indirect measurements, guaranteeing the successful implementation of a wide range of quantum information protocols.

Implementation of QST for discrete- and continuous-variable quantum systems usually is different owing to their different dimensions. For instance, in optical mode tomography, if the set of modes is discrete (e.g., in the case of polarization qubits), the photon-counting method \cite{2004LNP649P, RevModPhys.81.299} is used for characterization.
However, if the distribution of light particles over electromagnetic modes is described by a continuous degree of freedom (i.e., for continuous-variable QST), the homodyne tomography method \cite{RevModPhys.81.299} is employed. Irrespective of whether the system is discrete or continuous when considering their multiple cases, QST increases in complexity, raising their dependence on the experimental apparatus used. This remains a considerable challenge.

Mediated by quantum field theory (QFT), QIT has been extended to fit within the relativistic framework \cite{Mann_2012, RevModPhys.76.93, Hu_2012}, including the relativistic quantum information process \cite{Benincasa_2014, PhysRevLett.110.160501, doi:10.1142/S0219749906001736}, relativistic quantum metrology \cite{PhysRevD.89.065028, RQM1, PhysRevD.103.125025, RQM2, RQM3, FENG2022136992, Patterson:2022ewy}, fundamental tests of relativistic physics 
\cite{analogue-gravity, Tian, PhysRevA.90.052113, PhysRevD.105.066011, PhysRevD.105.125011, PhysRevD.103.085014, PhysRevLett.116.061301, PhysRevD.106.L061701, Jing}, and relativistic quantum information tasks \cite{PhysRevLett.91.180404, PhysRevLett.95.120404, PhysRevLett.110.113602, PhysRevD.91.084063, PhysRevD.101.085006, PhysRevA.106.032432}. This generalization confers theoretical completeness to QIT while heightening the necessary requirement for developing fast and efficient QST of quantum field systems.  However, the prospect of generalizing the traditional optics and oscillator state tomography techniques, such as homodyne measurements \cite{RevModPhys.81.299}, to processes involving quantum fields in a straightforward manner is demanding due to the following reasons: \emph{(1) the free field modes are spatially unlocalized and possess infinite degrees of freedom, and (2) “idealized measurements” are described by a collection $\{M_m\}$ of measurement operators, satisfying $\sum_mM^+_mM_m=\mathbb{I}$, $M_m=M^+_m$, and $M_mM_{m^\prime}=\delta_{m, m^\prime}M_m$. Upon measuring the system state $|\psi\rangle$, the probability of obtaining the result $m$ is given by $p(m)=\langle\psi|M^+_mM_m|\psi\rangle=\langle\psi|M_m|\psi\rangle$, and after the measurement, the state of the system is $M_m|\psi\rangle/\sqrt{\langle\psi|M^+_mM_m|\psi\rangle}=M_m|\psi\rangle/\sqrt{\langle\psi|M_m|\psi\rangle}$, which, in QFT, is physically impossible without violating causality \cite{Sorkin:1993gg, PhysRevA.64.052309, PhysRevD.104.025012, PhysRevD.103.025017, PhysRevD.105.025003}}. These challenges may prevent us from implementing QIT involving quantum fields. An approach to address these issues is by interacting with quantum fields through localized coupling with other systems, such as atoms or particle detectors \cite{Probe1, PhysRevD.97.105026, PhysRevD.98.105011, PhysRevD.104.085014, PhysRevD.98.105011, PhysRevD.105.065003}, allowing well-defined and physically meaningful measurements on quantum fields to be performed.
Nevertheless, the challenge of building effective measurements on dark quantum field systems, essential for realizing their complete QST and enabling the reliable implementation of QIT in the relativistic setting of QFT, continues to be elusive.

A Ramsey interferometry-based definition of work distributions in QFT has been proposed \cite{PhysRevLett.122.240604, PhysRevA.102.052219}. Moreover, based on a similar scheme, reconstructing the Wigner characteristic function of quantum states of an infinite dimensional system, such as the motional oscillation of a harmonic oscillator \cite{PhysRevLett.122.110406, PhysRevA.104.012421}, has been investigated. Implementation via a trapped-ion system was presented \cite{PhysRevLett.125.043602}.
Encouraged by this idea, herein, a Ramsey scheme--based direct characteristic-function tomography of quantum fields is proposed without the requirement of any direct control and the associated “idealized measurements” on quantum fields, that are always ill-defined in QFT. Specifically, operation on quantum fields is performed through a locally coupled particle detector: Detector state rotations are appropriately selected in conjunction with detector state-dependent displacements, the expectation value of the real and imaginary parts of the displacement operator is mapped to the detector states; finally, these states are read out to reconstruct the symmetric characteristic function. Its applicability is illustrated via its implementation in the Kubo--Martin--Schwinger (KMS) states and squeezed states of a quantum scalar field. In addition, its general application to analog quantum fields in analogue-gravity systems, such as Bose--Einstein condensates (BEC), is examined within state-of-the-art experimental capabilities.

This paper is presented as follows. Sec. \ref{section1} reviews the characteristic function that can be employed to completely describe the quantum state of the continuous-variable system. Sec. \ref{section2} presents a comprehensive investigation of how to exploit an auxiliary qubit to implement a direct readout of the symmetric characteristic function of the quantum states of quantum fields through the qubit--field interaction. The proposed approach is applied to the Kubo--Martin--Schwinger (thermal) states and squeezed states of a quantum scalar field. Sec. \ref{section3} discusses the general application of the proposed technologies. Finally, Sec. \ref{section4} presents the conclusions.

%%%%%%%%%%%%%%%%%%%%%%%%%%%%%%%%%%%%%%%%%%%%%%%%%%%%%%%%%%%%
\section{Background}\label{section1}

Besides the state density matrix, a complete description of the quantum state of the continuous-variable system can also be provided by the characteristic function of quasiprobability distribution \cite{barnett2002methods}. For an arbitrary quantum state $\hat{\rho}$, its equivalent symmetric characteristic function is defined as
\begin{eqnarray}\label{CF}
\chi(\boldsymbol{\xi})=\mathrm{Tr}\big[\hat{\rho}\hat{D}(\boldsymbol{\xi})\big]=\mathrm{Tr}\bigg[\hat{\rho}\bigotimes_{i=1}^n\hat{D}_i(\xi_i)\bigg],
\end{eqnarray}
where $\boldsymbol{\xi}=(\xi_1, \dots, \xi_n)^T$ is the column vector with $\xi_i\in\mathbb{C}$ and $\hat{D}(\xi_i)=\exp\{\xi_i\hat{a}^\dagger-\xi_i^\ast\hat{a}_i\}$ is the single-mode displacement operator. The quasiprobability distribution can be obtained from the characteristic function \eqref{CF} via Fourier transform
\cite{ferraro2005gaussian, RevModPhys.84.621}
\begin{eqnarray}\label{QD}
W(\boldsymbol{\alpha})=\int_{\mathbb{R}^{2n}}\frac{d^{2n}\boldsymbol{\xi}}{(2\pi)^{2n}}
\exp\big\{\boldsymbol{\xi}^\dagger\boldsymbol{\alpha}+\boldsymbol{\alpha}^\dagger\boldsymbol{\xi}\big\}\chi(\boldsymbol{\xi}),
\end{eqnarray}
where the continuous variables $\boldsymbol{\alpha}\in\mathbb{R}^{2n}$ are eigenvalues of quadrature operators $\hat{\boldsymbol{\alpha}}=(\hat{x}_1, \hat{p}_1, \dots, \hat{x}_n, \hat{p}_n)^T$, with $\hat{x}_i$ being the position operator and $\hat{p}_i$ being the momentum operator. Notably, an arbitrary quantum state $\hat{\rho}$ of a $n$-mode bosonic system is equivalent to a Wigner function $W(\boldsymbol{\alpha})$ defined over a $2n$-dimensional phase space. In addition, expectation values of powers of creation and destruction operators \cite{barnett2002methods, ferraro2005gaussian} are given by derivatives of the characteristic function around the origin:
\begin{eqnarray}\label{operators-expectation}
\mathrm{Tr}\bigg[\hat{\rho}\bigg[\big(\hat{a}_i^\dagger\big)^p\hat{a}_l^q\bigg]_\text{S}\bigg]=(-1)^q\frac{\partial^{p+q}}{\partial\xi^p_i\partial\xi^{\ast\,q}_l}\chi(\boldsymbol{\xi})\bigg|_{\boldsymbol{\xi}=0}.
\end{eqnarray}
Here, the subscript S denotes symmetric operator ordering \cite{barnett2002methods, ferraro2005gaussian}. Obtaining similar information requires integration of the Wigner function over the complete phase space.

\begin{figure}
\centering
\includegraphics[width=0.42\textwidth]{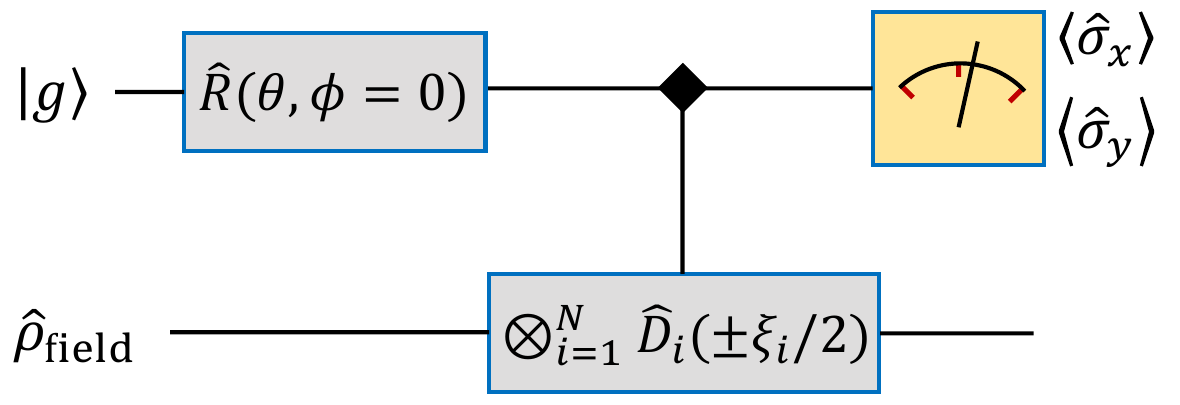}
\caption{Characteristic-function readout of the quantum field state $\hat{\rho}_\text{field}$ through the Ramsey interferometry scheme.
A rotation $\hat{R}(\theta, \phi=0)$ is applied on the qubit’s initial state $|g\rangle$; subsequently, the quantum fields are displaced by $\otimes^N_{i=1}\hat{D}_i(\pm\xi_i/2)$, where the sign relies on the excitation/ground state of the qubit; finally, the auxiliary qubit states are read out.
}\label{fig1}
\end{figure}

%%%%%%%%%%%%%%%%%%%%%%%%%%%%%%%%%%%%%%%%%%%%%%%%%%%%%%%%%%%%
\section{Characteristic-function reconstruction scheme of quantum fields} \label{section2}

This section outlines the Ramsey scheme--based direct reconstruction of the characteristic function of the quantum states of quantum fields. Indeed, this proposed scheme is well-defined for QFT despite the impossibility of direct control and the associated idealized measurements on quantum fields.

\subsection{Characteristic-function reconstruction scheme}

Briefly, the quantum fields of interest are coupled to an auxiliary qubit, which exerts a state-dependent potential on the quantum fields.
If the qubit is prepared in a superposition of ground and excited states, the dynamical process may transfer the data about the characteristic function of the quantum states of quantum fields to the qubit state. Fig. \ref{fig1} illustrates the steps as follows:

1. The auxiliary qubit and the quantum field system are initially prepared in the product state $|g\rangle\langle\,g|\otimes\hat{\rho}_\text{field}$.

2. A rotation operation $\hat{R}(\theta, \phi)=\cos(\theta/2)\mathbb{I}-i\sin(\theta/2)[\cos(\phi)\sigma_x+\sin(\phi)\sigma_y]$ is applied on the qubit.

3. The field system and the qubit evolve unitarily based on 
$
\hat{U}_\text{int}(T)=|0\rangle\langle0|\otimes\hat{D}(-\boldsymbol{\xi}(T)/2)+|1\rangle\langle1|\otimes\hat{D}(\boldsymbol{\xi}(T)/2),
$
where $\hat{D}(\pm\boldsymbol{\xi}(T)/2)=\otimes^N_{i=1}\hat{D}_i(\pm\xi_i(T)/2)$, with $T$ being the evolution time interval.

4. Finally, the reduced state of the auxiliary qubit is given by
$\hat{\rho}(T)=\frac{1}{2}[1-\cos(\theta)\sigma_z+\sin(\theta)(\Im[\chi(\boldsymbol{\xi})]\sigma_x+\Re[\chi(\boldsymbol{\xi})]\sigma_y)]$.
By measuring the Pauli operators $\sigma_x$ and $\sigma_y$ with respect to the reduced state of the auxiliary qubit $\rho(T)$ and iterating this process over different values of $\boldsymbol{\xi}$, the characteristic function $\chi(\boldsymbol{\xi})$ is reconstructed.

To illustrate the implementation of the aforementioned proposal, let us take a concrete example. Measuring the characteristic function of a massive scalar quantum field $\hat{\phi}(t, \bx)$ in $(n+1)$-dimensional Minkowski spacetime is attempted without loss of generality. The field can be written in terms of plane-wave modes as
$\hat{\phi}(t, \bx)=\int\frac{d^n\bk}{\sqrt{2(2\pi)^n\omega_\bk}}(\hat{a}_\bk\,e^{i\mathsf{k}\cdot\mathsf{x}}+\hat{a}^\dagger_\bk\,e^{-i\mathsf{k}\cdot\mathsf{x}})$,
in the Heisenberg picture.
Here, $\mathsf{k}\cdot\mathsf{x}=\bk\cdot\bx-\omega_\bk\,t$, $\omega_\bk=\sqrt{m^2+\bk^2}$, and the creation and destruction operators satisfy
$[\hat{a}_\bk, \hat{a}^\dagger_{\bk^\prime}]=\delta^n(\bk-\bk^\prime)$. For many purposes, it becomes more convenient to restrict the field to the interior of a spacelike $n$-torus of side $L$ (i.e., select periodic boundary conditions) \cite{birrell1984quantum}, following which the field can be rewritten as
\begin{eqnarray}
\hat{\phi}(t, \bx)=\sum_\bk(2L^{n}\omega_\bk)^{-\frac{1}{2}}(\hat{a}_\bk\,e^{i\mathsf{k}\cdot\mathsf{x}}+\hat{a}^\dagger_\bk\,e^{-i\mathrm{k}\cdot\mathrm{x}}),
\end{eqnarray}
where $k_i=2\pi\,j_i/L$ with $j_i=0, \pm1, \pm2, \dots$ and $i=1, \dots, n$. The free Hamiltonian of the field takes the form $\hat{H}_{\hat{\phi}}=\sum_\bk\omega_\bk\hat{a}^\dagger_\bk\hat{a}_\bk$.
Furthermore, the auxiliary qubit, possessing a free Hamiltonian $\hat{H}_q=\frac{1}{2}\omega_0\sigma_z$, is assumed to interact with the scalar quantum field through
\begin{eqnarray}\label{Interaction}
\nonumber
\hat{H}_\text{int}&=&\lambda\eta(t)\sigma_z\int\,d^n\mathbf{x}F(\mathbf{x})\hat{\phi}(\mathbf{x})
\\
&=&|g\rangle\langle\,g|\hat{H}_g+|e\rangle\langle\,e|\hat{H}_e.
\end{eqnarray}
Here, $\lambda$ denotes the coupling strength, $\eta(t)$ represents the switching function controlling the time dependence of the coupling, and $F(\mathbf{x})$ refers to the spatial smearing function that models the localization of the auxiliary qubit. The interaction shown in Eq. \eqref{Interaction} may provide the effective qubit-state-dependent potentials $\hat{H}_g$ and $\hat{H}_e$ for the quantum field. In the interaction picture with respect to $\hat{H}_q$, the dynamics of the combined system is governed by
\begin{eqnarray}\label{H1}
\hat{H}=|g\rangle\langle\,g|\hat{V}_g+|e\rangle\langle\,e|\hat{V}_e,
\end{eqnarray}
where $\hat{V}_g=\hat{H}_{\hat{\phi}}+\hat{H}_g$ and $\hat{V}_e=\hat{H}_{\hat{\phi}}+\hat{H}_e$ act on the field.
In what follows, this dynamics will be used to obtain information on the quantum field state.

To this end, a rotation is applied on the auxiliary qubit initialized in the ground state, following which the qubit is rotated to the superposition state $|\psi(0)\rangle_\text{q}=\cos\frac{\theta}{2}|g\rangle-i\sin\frac{\theta}{2}|e\rangle$. Subsequently, the state of the whole system $|\psi(0)\rangle_\text{q}{_\text{q}}\langle\psi(0)|\otimes\hat{\rho}_\text{field}$ evolves unitarily under the Hamiltonian \eqref{H1} and becomes entangled. This dynamics enables measurements on the auxiliary qubit to connect with quantities of the quantum field \cite{ PhysRevD.105.065003}. Moreover, to realize the controllable displacement and obtain more information, a series of pulses are actually utilized to manipulate the auxiliary qubit. Specifically, $2N$ $\pi$ pulses are applied, each being separated by the free-evolution time $\tau$.
Assuming the total evolution time $T=2N\tau$, the final state of the auxiliary qubit is given by (see the Supplemental Material \cite{Supplemental-Material} for details)
\begin{eqnarray}\label{final state}
\nonumber
\hat{\rho}_\text{q}(T)&=&\mathrm{Tr}_\text{field}\big[\hat{U}(T)|\psi(0)\rangle_\text{q}{_\text{q}}\langle\psi(0)|\otimes\hat{\rho}_\text{field}\hat{U}^\dagger(T)\big]
\\        \nonumber
&=&\frac{1}{2}\big[1-\cos(\theta)\sigma_z+\sin(\theta)(\Im[\chi(\boldsymbol{\xi}_\bk)]\sigma_x
\\
&&+\Re[\chi(\boldsymbol{\xi}_\bk)]\sigma_y)\big].
\end{eqnarray}
From Eq. \eqref{final state}, the field characteristic function is encoded into the evolved state of the auxiliary qubit as long as $\sin\theta\neq0$. Performing proper measurements, $\sigma_x$ and $\sigma_y$, on the auxiliary qubit state \eqref{final state}, $\langle\sigma_y\rangle+i\langle\sigma_x\rangle=\sin\theta(\Re[\chi(\boldsymbol{\xi}_\bk)]+i\Im[\chi(\boldsymbol{\xi}_\bk)])=\sin\theta\chi(\boldsymbol{\xi}_\bk)$. Thus, the rotation operator on the auxiliary qubit introduced above should satisfy $\sin\theta\neq0$ to guarantee the success of the protocol proposed in this work. Note that if $\theta=\pi/2$ is taken, after the proper measurements, one can directly obtain $\langle\sigma_y\rangle+i\langle\sigma_x\rangle=\mathrm{Tr}_\text{field}\big[\hat{\rho}_\text{field}\exp\big\{\sum_\bk\big[\xi_\bk\hat{a}^\dagger_\bk-\xi^\ast_\bk\hat{a}_\bk\big]\big\}\big]=\chi(\boldsymbol{\xi}_\bk)$,
which provides a direct link between the characteristic function $\chi(\boldsymbol{\xi}_\bk)$ and the measure data without the need for any integral transform of the latter. Here, the displacement $\xi_\bk$ of mode $\bk$ is given by
\begin{eqnarray}\label{xi}
\nonumber
\xi_\bk&=&-4\frac{\lambda\widetilde{\eta}_\bk(\tau)\widetilde{F}^\ast(\bk)}{\omega_\bk\tau}\sin(N\omega_\bk\tau)\tan\bigg(\frac{\omega_\bk\tau}{2}\bigg)
e^{iN\omega_\bk\tau},
\\
\end{eqnarray}
where $\widetilde{F}(\bk)=\int\,d^n\bx\,F(\bx)e^{i\bk\cdot\bx}$ denotes the Fourier transform of the spatial smearing function $F(\bx)$ and $\widetilde{\eta}_\bk(\tau)=\int^\tau_0\frac{\eta(s)}{\sqrt{2L^{n}\omega_\bk}}ds$. Notably, the characteristic function contains all the information regarding the initial density operator $\hat{\rho}_\text{field}$ of the quantum field and can be used to obtain the Wigner function through the complex Fourier transform \eqref{QD}. Moreover, the set of displacement operators $\hat{D}(\boldsymbol{\xi})$ is complete, signifying that it allows for the representation of any operator $\hat{O}$ as $\hat{O}=\int_{\mathbb{R}^{2n}}\frac{d^{2n}\boldsymbol{\xi}}{\pi^{n}}\text{Tr}[\hat{O}\hat{D}(\boldsymbol{\xi})]\hat{D}^\dagger(\boldsymbol{\xi})$ \cite{ferraro2005gaussian}. Thus, by assuming $\hat{O}=\hat{\rho}_\text{field}$, the density operator itself can be exactly reconstructed with the complete knowledge of the characteristic function.

Here, it is stressed that the proposed scheme only involves interactions with a low-dimensional ancilla and provides an indirect way to locally measure the quantum field. It does not require any idealized measurements on the quantum field, for instance, the projection operation of quadratures, which are always ill-defined in the QFT.
Moreover, the scheme involves only manipulating the auxiliary qubit while without any manipulation of the field such as a displacement operation prior to the measurement procedure, or a control of the coupling strength. Hence, the proposed scheme is not only well defined in QFT but also experimentally feasible in principle, despite direct manipulation on the quantum field typically being difficult to realize.

\subsection{Characteristic-function tomography of the finite-temperature KMS state and the squeezed state}

For the field in a finite-temperature KMS state $\hat{\rho}_\beta$ (with inverse KMS temperature $\beta$) \cite{doi:10.1143/JPSJ.12.570, PhysRev.115.1342} or
a squeezed state $|\boldsymbol{\zeta}\rangle=\hat{S}_{\boldsymbol{\zeta}}|0\rangle=\exp\big[\frac{1}{2}\sum_\bk\big(\zeta^\ast_\bk\hat{a}^2_\bk-\zeta_\bk\hat{a}^{\dagger2}_\bk\big)\big]|0\rangle$ \cite{PhysRevD.98.085007}, their characteristic functions can be derived (\cite{Supplemental-Material})
\begin{eqnarray}\label{MTCF}
\chi^{\beta}(\boldsymbol{\xi}_\bk)=\exp\bigg[-\frac{1}{2}\sum_\bk|\xi_\bk|^2\frac{e^{\beta\omega_\bk}+1}{e^{\beta\omega_\bk}-1}\bigg],
\end{eqnarray}
and
\begin{eqnarray}\label{MSCF}
\nonumber
\chi^{\zeta}(\boldsymbol{\xi}_\bk)&=&\exp\bigg[-\frac{1}{2}\sum_\bk\big(\cosh[2\zeta_\bk]|\xi_\bk|^2-\sinh[2\zeta_\bk]
\\
&&\times\Re\big[e^{i\theta_\bk}\xi^{\ast2}_\bk\big]\big)\bigg].
\end{eqnarray}
In the limit $\beta\rightarrow\infty$ or $\zeta_\bk\rightarrow0$, that is, zero-temperature limit or infinitely small squeezing, Eqs. \eqref{MTCF} and \eqref{MSCF} represent the characteristic function of the vacuum state of the field.

\begin{figure}
\centering
\includegraphics[width=0.42\textwidth]{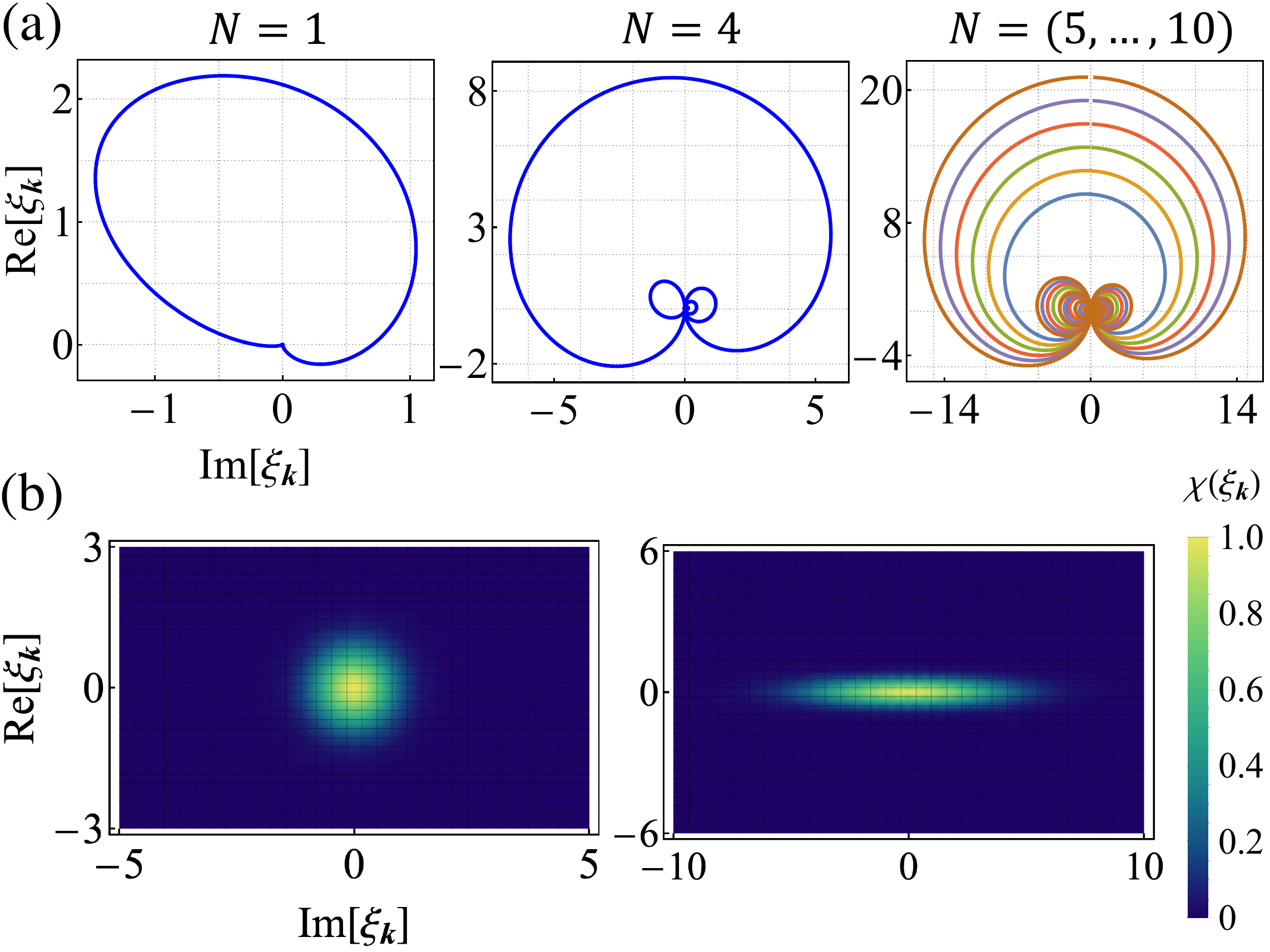}
\caption{ State reconstruction of a single-mode quantum field. (a) Reachable points $\xi_\bk$ (with unit $\omega_\bk/\lambda$) in reciprocal phase space for different numbers of pulse cycles: $N=1$ and $N=4$ in the first two panels and all points for $N= (5, \dots, 10)$ in the right panel. Here, the normalized spherical Gaussian centered at zero is taken as smearing function $F(x)=e^{-\frac{r^2}{2\sigma^2}}/(\sqrt{2\pi\sigma^2})^3$, with $\sigma=0.1/\omega_\bk$, and the switching function $\eta(s)=e^{\frac{-(s-T/2)^2}{T^2/72}}$ with $T=4/\omega_\bk$ is selected. In addition, the dimensionless parameter $\omega_\bk\,s$ is adopted in the plots. (b) Readouts of the characteristic functions for the KMS state \eqref{Thermal-C} with $n=1/(e^{\beta\omega_\bk}-1)=1$ (the left panel) and for the squeezed state \eqref{Squeezed-C} with $\zeta_\bk=1$ (the right panel).}\label{fig2a}\label{fig2}
\end{figure}

The proposed scheme remains applicable when considering an auxiliary qubit coupled to a single-mode quantum field \cite{PhysRevLett.121.071301, doi:10.1073/pnas.1807703115, PhysRevResearch.1.033027}. In such a case, the corresponding single-mode characteristic functions for the KMS state and squeezed state are, respectively, given by
\begin{eqnarray}\label{Thermal-C}
\chi^{\beta}(\xi_\bk)=\exp\bigg[-\frac{1}{2}|\xi_\bk|^2\frac{e^{\beta\omega_\bk}+1}{e^{\beta\omega_\bk}-1}\bigg],
\end{eqnarray}
and
\begin{eqnarray}\label{Squeezed-C}
\nonumber
\chi^{\zeta}(\xi_\bk)&=&\exp\bigg[-\frac{1}{2}\big(\cosh[2\zeta_\bk]|\xi_\bk|^2-\sinh[2\zeta_\bk]
\\
&&\times\Re\big[e^{i\theta_\bk}\xi^{\ast2}_\bk\big]\big)\bigg].
\end{eqnarray}
The displacement parameter $\xi_\bk$ exhibits periodicity with a period of $\tau=2\pi/\omega_\bk$ and attains a maximum value
\begin{eqnarray}
\xi_\bk=8\frac{\lambda\,N\widetilde{\eta}_\bk(\pi/\omega_\bk)\widetilde{F}^\ast(\bk)}{\pi},
\end{eqnarray}
when $\tau=\pi/\omega_\bk$. Notably, this maximal value scales linearly with the pulse number $N$. For each fixed pulse-sequence parameter $N$, the corresponding displacement parameter $\xi_\bk$ in the reciprocal phase space is a close curve as shown in Fig. \ref{fig2}. By varying $N$, the corresponding different closed curves for $\xi_\bk$ can be obtained, and different displacement operations on the field are realized. Along the manifold of curves, the characteristic function can be sampled, and Fig. \ref{fig2} illustrates the cases for the KMS state and squeezed state.
Performing the Fourier transform given in \eqref{QD}, the corresponding quasiprobability distributions---Wigner functions for the target states---can be obtained.
The corresponding characteristic functions of both the KMS state and squeezed state are mostly centered around the origin, at which they possess the maximum $\chi(0)=1$, as shown in Fig. \ref{fig2}. This property proves convenient for experiments because of the close proximity of the density of reachable points $\xi_\bk$ to the origin.

\begin{figure}
\centering
\includegraphics[width=0.42\textwidth]{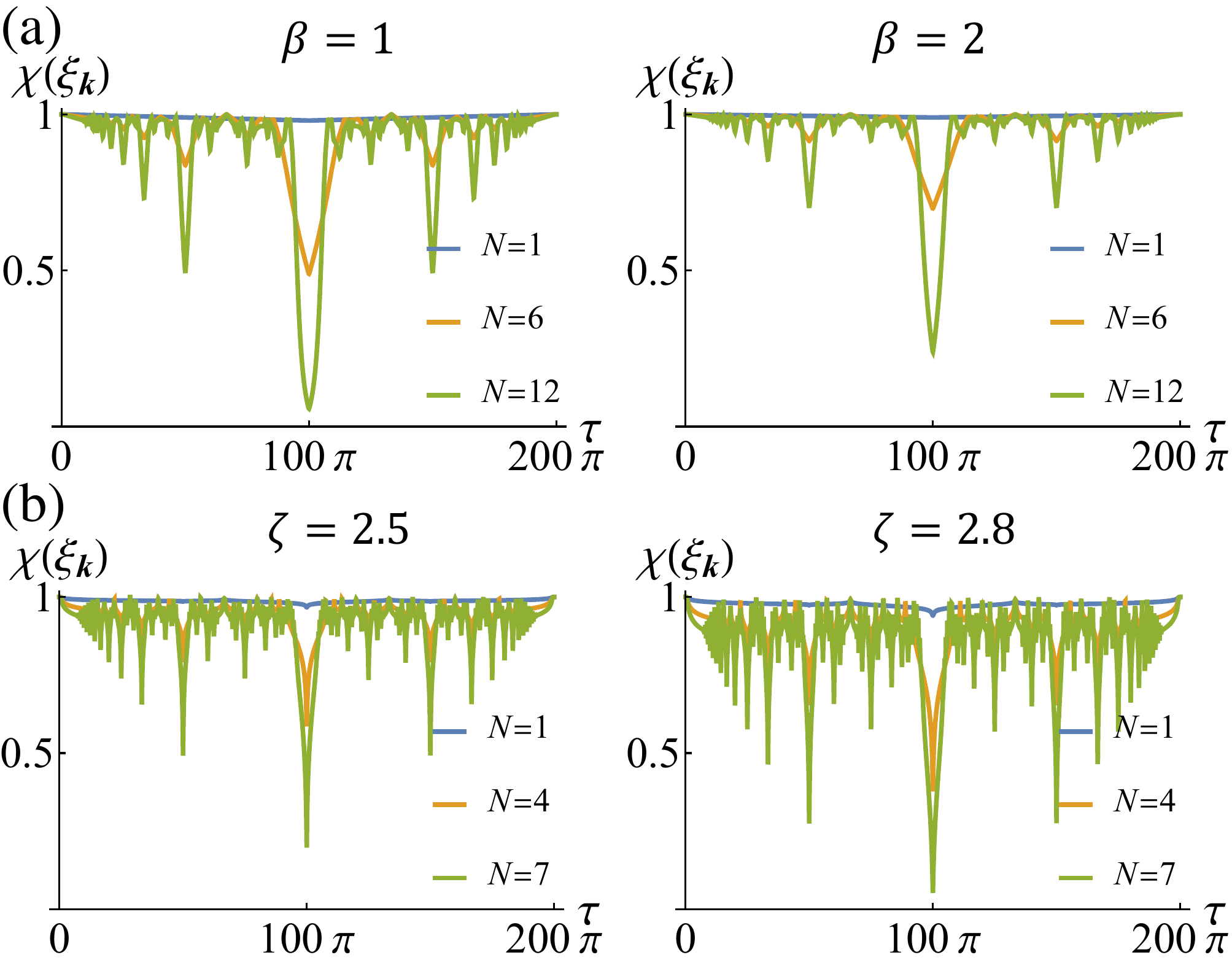}
\caption{Reconstruction of the thermal state in \eqref{MTCF} (a) and the squeezed state in \eqref{MSCF} (b). Here, the normalized spherical Gaussian centered at zero is taken as smearing function $F(x)=e^{-\frac{r^2}{2\sigma^2}}/(\sqrt{2\pi\sigma^2})^3$, with $\sigma=1$, and the switching function $\eta(s)=1$ is selected for simplification. The coupling constant $\lambda=0.01$ is assumed.}\label{fig3}
\end{figure}

In the case of multimode, the characteristic functions in Eqs.\eqref{MTCF} and \eqref{MSCF} contain contributions from all the modes.
Fig. \ref{fig3} reveals that the characteristic functions for the multimode KMS state and squeezed state are plotted via the free-evolution time, $\tau$, with different pulse numbers, $N$, which are the main controllable parameters of $\boldsymbol{\xi}_\bk$. Note that, in this plot, the switching function $\eta(s)=1$ is selected for simplicity. Although this choice may affect $\boldsymbol{\xi}_\bk$ at different time $\tau$ and with different pulse numbers $N$, it would not change the map between the characteristic function $\chi(\boldsymbol{\xi}_\bk)$ and the variables $\boldsymbol{\xi}_\bk$ shown in Eqs. \eqref{MTCF} and \eqref{MSCF}. In addition, to derive the quasiprobability distribution $W(\boldsymbol{\alpha})$ shown in Eq. \eqref{QD}, the Fourier transform of the characteristic function $\chi(\boldsymbol{\xi}_\bk)$ is required. This implies that, in order to derive the quasiprobability distribution, we need information about the distribution of $\chi(\boldsymbol{\xi}_\bk)$ over $\boldsymbol{\xi}_\bk$ in theory regardless of the functional form of $\boldsymbol{\xi}_\bk$. In this sense, the selection of the switching function would not change this map and, thus, would not change the relevant results concerned.

Note that the characteristic function $\chi(\boldsymbol{\xi}_\bk)^\ast=\chi(\boldsymbol{-\xi}_\bk)$ is Hermitian; thus, any half of the complex space covered by $\boldsymbol{\xi}_\bk$ is sufficient for a complete measurement. This property means that the scheme proposed in this study to measure the characteristic function may reduce the magnitude of required readouts. Usually, complete reconstruction of the entire $\chi(\boldsymbol{\xi}_\bk)$ for an infinite dimensional system is complex and impractical. However, several interesting properties can be accessed given only a finite collection of $\chi(\boldsymbol{\xi}_\bk)$ values. For instance, one can directly estimate the nonclassicality of quantum field states from a finite collection of characteristic function values through the nonclassicality criteria in Ref. \cite{PhysRevLett.106.010403}, so do the estimation of entanglement with lower bounds for entanglement measurements. Additionally, from Eq. \eqref{operators-expectation}, one can estimate the expectation values of powers of creation and destruction operators only with a finite set of characteristic function values, that is, measuring them only in the vicinity of the $\boldsymbol{\xi}_\bk=0$ regime. This property allows estimating the correlation function between different modes, which are of broad interest---e.g., in many-body model simulation.

Our interest is the distribution of the characteristic function $\chi(\boldsymbol{\xi}_\bk)$ over the parameter space $\boldsymbol{\xi}_\bk$ as shown, for instance, in Fig. \ref{fig2} (b). To characterize the target characteristic function as completely as possible, it is assumed that all the corresponding values of the characteristic function should be known to at least $N$ points in the $\boldsymbol{\xi}_\bk$ space around the origin.
Through the $\langle\sigma_y\rangle+i\langle\sigma_x\rangle=\chi(\boldsymbol{\xi}_\bk)$ given above, for each sampled point measurement, the error of the characteristic function $\chi_j$ is $\Delta_{\chi_j}=\sqrt{\Delta^2_{\sigma_{x_j}}+\Delta^2_{\sigma_{y_j}}}$, with $j={1, 2, \dots, N}$. Here, $\Delta_{\sigma_{x_j}}$ and $\Delta_{\sigma_{y_j}}$ refer to the errors for measuring $\sigma_x$ and $\sigma_y$, respectively, which satisfy $\Delta_{\sigma_{x_j}}\sim1/\sqrt{M}$ and $\Delta_{\sigma_{y_j}}\sim1/\sqrt{M}$, with the measurement number $M$. Thus, to achieve a target error $\Delta_{\chi_j}$, the needed measurement number is $M\sim1/\Delta^2_{\chi_j}$.

%%%%%%%%%%%%%%%%%%%%%%%%%%%%%%%%%%%%%%%%%%%%%%%%%%%%%%%%%%%
\section{Application to the analogue gravity system}\label{section3}
Analogue gravity is a research programme that investigates analogues of general relativistic gravitational fields within other physical systems, such as BEC \cite{analogue-gravity}, superconducting circuits \cite{RevModPhys.84.1}, trapped ions \cite{RevModPhys.86.153, Johanning_2009} and so on, with the objective of gaining new insights into their corresponding problems. These quantum simulators are termed analogue-gravity systems. Recently, they have been extensively applied to the simulation of relativistic quantum effects, which are usually inaccessible for quantum fields in real relativistic settings.
In such proposed experiments, numerous different quantum states of analogue quantum fields have been arbitrarily assumed and exploited. However, the prerequisite to all the experimental manipulation---characterizing quantum states—is deliberately circumvented due to unknown relevant technologies. The scheme proposed in this work may be valid for solving such issues.

Herein, the proposed scheme is applied to a BEC system. Usually, the density fluctuations in BEC are modeled as analogue quantum fields, and an impurity immersed in the BEC is considered as a detector \cite{PhysRevLett.91.240407, PhysRevLett.94.040404, PhysRevD.69.064021, PhysRevLett.118.045301, PhysRevResearch.2.042009, PhysRevD.103.085014, PhysRevD.106.L061701}. The impurity is coupled to the density fluctuations \cite{PhysRevLett.91.240407, PhysRevLett.94.040404, PhysRevD.69.064021} and, thus, can work as the auxiliary qubit in this situation.
The objective is to implement the quantum-state reconstruction of density fluctuations in the BEC here.
Similar to Refs. \cite{PhysRevLett.118.045301, PhysRevResearch.2.042009, PhysRevD.103.085014, PhysRevD.106.L061701}, the impurity consisting of a two-level atom with internal Hamiltonian $\hat{H}_q=\omega_0|e\rangle\langle\,e|$ interacts with the BEC via $\hat{H}_\text{int}=\sum_{s=g, e}g_s\hat{\rho}(\mathbf{r}_A)|s\rangle\langle\,s|$, where the Bose gas density reads $\hat{\rho}(\mathbf{r}_A)=\rho_0+\delta\hat{\rho}(\mathbf{r}_A)$ with $\rho_0$ being a constant and the density fluctuations $\delta\hat{\rho}(\mathbf{r}_A)=\sqrt{\rho_0}\int\big[d^3k/(2\pi)^3\big]\big(u_k+v_k\big)\big[e^{i\bk\cdot\,\br}b_\bk+e^{-i\bk\cdot\,\br}b^+_\bk\big]$. In addition, the free Hamiltonian of the density fluctuations is $\hat{H}_\rho=\int\,d^3k\omega_\bk\,b^+_\bk\,b_\bk$. In the interaction picture with respect to $\hat{H}_q$, the same form of Hamiltonian in \eqref{H1} can be obtained.
Here, the coupling strength is determined by $g_s$, and the term $u_k+v_k=\sqrt{E_k/\omega_k}$ related to the Bogoliubov coefficients \cite{RevModPhys.77.187} is similar to the spatial smearing function after Fourier transformation. Repeating the same analysis above, in principle, the symmetric characteristic function of the density fluctuations, that is, analogue quantum fields in the quantum simulator, can be reconstructed.

%%%%%%%%%%%%%%%%%%%%%%%%%%%%%%%%%%%%%%%%%%%%%%%%%%%%%%%%%%%
\section{Conclusions}\label{section4}

In this work, we propose a general scheme for the quantum-state reconstruction of quantum fields, which usually is inaccessible for direct control and measurements.
This approach only relies on the quantum control and measurements of a localized auxiliary qubit coupled to the quantum fields without any manipulations of the target quantum fields. We characterize this technique by application to the KMS state and squeezed state of a quantum scalar field, and moreover, its application to the BEC, an analogue-gravity system, is discussed. This proposed scheme solves the problem of quantum-state reconstruction of quantum fields and, thus, facilitates the development of the modern implementation of quantum information protocols involving quantum fields.

Moreover, our scheme can be applied to the quantum-state reconstruction of quantum fields in various spacetime backgrounds, e.g., strong gravitational field and relativistic framework of motions, where the direct physical measurements and manipulations on quantum fields remain unknown.
This generalization may provide an essential tool to understand and optimize the control of quantum fields for relativistic quantum information applications, especially for exploring the interplay of gravity and quantum mechanics.

%%%%%%%%%%%%%%%%%%%%%%%%%%%%%%%%%%%%%%%%%%%%%%%%%%%%%%%%%
~~~~~
\begin{acknowledgments}
This work was supported by the National Key R\&D Program of China (Grant No. 2018YFA0306600), and Anhui Initiative in Quantum Information Technologies (Grant No. AHY050000).  
ZT was supported by the National Natural Science Foundation of 
China under Grant No. 11905218, and the CAS Key Laboratory for Research in Galaxies and Cosmology, Chinese Academy of Science (No. 18010203). 
\end{acknowledgments}

%\newpage

%\bibliography{CF-Tomography-QF}
%

%%%%%%%%%%%%%%%%%%%%%%%%%%%%%%%%%%%%%%%%%%
%%%%%%%%%%%%%%%%%%%%%%%%%%%%%%%%%%%%%%%%%%
%%%%%%%%%%%%%%%%%%%%%%%%%%%%%%%%%%%%%%%%%%
%%%%%%%%%%%%%%%%%%%%%%%%%%%%%%%%%%%%%%%%%%
%%%%%%%%%%%%%%%%%%%%%%%%%%%%%%%%%%%%%%%%%%

\vspace{1.5cm}

%\newpage
%==========================================================================================================
%==========================================================================================================
%==================================== Supplemental Material ====================================================
%==========================================================================================================
%==========================================================================================================
\pagebreak
\clearpage
\widetext

\begin{center}
\textbf{\large Supplementary Material}
\end{center}

\setcounter{equation}{0}
\setcounter{section}{0}
\setcounter{page}{1}
\makeatletter
\renewcommand{\theequation}{S\arabic{equation}}

\section{The qubit-state-dependent operator on quantum field}
With the Hamiltonian \eqref{H1} given in the main text, the time evolution of both the auxiliary qubit and the quantum field is determined by the operator
\begin{eqnarray}
\hat{U}(\tau)=|g\rangle\langle\,g|\exp\bigg[-i\int^\tau_0\,ds\hat{V}_g\bigg]+|e\rangle\langle\,e|\exp\bigg[-i\int^\tau_0\,ds\hat{V}_e\bigg],
\end{eqnarray}  
where $\tau$ is the evolution time interval.
It means the time evolution of the quantum field depends on the state of the auxiliary qubit. With this element, 
we can construct different probe-state-dependent unitary evolution operators on the quantum field, through a series of pulses on the qubit. Then the operators constructed may allow us to obtain appreciably more information of quantum field during the measurement process, as shown in the main text. For example, if we apply the pulse sequence $[\tau-\pi-\tau-\pi]$, i.e., an evolution time $\tau$, a $\pi$ pulse, and a subsequent second evolution time $\tau$ followed by a final $\pi$ pulse, 
the two qubit-state-dependent unitary evolution operators on quantum field for such a segment respectively read
\begin{eqnarray}
\hat{u}_g=\exp\bigg[-i\int^\tau_0\,ds\hat{V}_e\bigg]\exp\bigg[-i\int^\tau_0\,ds\hat{V}_g\bigg]~~~\text{and}~~~\hat{u}_e=\exp\bigg[-i\int^\tau_0\,ds\hat{V}_g\bigg]\exp\bigg[-i\int^\tau_0\,ds\hat{V}_e\bigg].
\end{eqnarray}
Here the subscript, $g$/$e$, denotes the ground/excitation state dependent.
The two above operators can further be rewritten as  
\begin{eqnarray}
\nonumber
\hat{u}_g&=&\exp\bigg[-i\int^\tau_0\,ds\bigg(\sum_\bk\omega_\bk\hat{a}^\dagger_\bk\hat{a}_\bk+\lambda\eta(s)\int\,d^n\bx\,F(\bx)\hat{\phi}(\bx)\bigg)\bigg]\exp\bigg[-i\int^\tau_0\,ds\bigg(\sum_\bk\omega_\bk\hat{a}^\dagger_\bk\hat{a}_\bk-\lambda\eta(s)\int\,d^n\bx\,F(\bx)\hat{\phi}(\bx)\bigg)\bigg]
\\   \nonumber
&=&\exp\bigg[-i\sum_\bk\bigg(\omega_\bk\tau\hat{a}^\dagger_\bk\hat{a}_\bk+\lambda\widetilde{\eta}_\bk(\tau)\big(\widetilde{F}(\bk)\hat{a}_\bk+\widetilde{F}^\ast(\bk)\hat{a}^\dagger_\bk\big)\bigg)\bigg]
\exp\bigg[-i\sum_\bk\bigg(\omega_\bk\tau\hat{a}^\dagger_\bk\hat{a}_\bk-\lambda\widetilde{\eta}_\bk(\tau)\big(\widetilde{F}(\bk)\hat{a}_\bk+\widetilde{F}^\ast(\bk)\hat{a}^\dagger_\bk\big)\bigg)\bigg]
\\   \nonumber
&=&\exp[i\varphi_1]\hat{D}^\dagger(\boldsymbol{\epsilon}_\bk)\exp\bigg[-\sum_{\bk}\,2i\omega_{\bk}\tau\hat{a}^\dagger_\bk\hat{a}_\bk\bigg]
\hat{D}\big(2\boldsymbol{\epsilon}_\bk\,e^{i\boldsymbol{\omega}_\bk\tau}\big)\hat{D}^\dagger(\boldsymbol{\epsilon}_\bk),
\\
\hat{u}_e&=&\exp[-i\varphi_1]\hat{D}(\boldsymbol{\epsilon}_\bk)\exp\bigg[-\sum_{\bk}\,2i\omega_{\bk}\tau\hat{a}^\dagger_\bk\hat{a}_\bk\bigg]
\hat{D}^\dagger\big(2\boldsymbol{\epsilon}_\bk\,e^{i\boldsymbol{\omega}_\bk\tau}\big)\hat{D}(\boldsymbol{\epsilon}_\bk).
\end{eqnarray}
where $\widetilde{F}(\bk)=\int\,d^n\bx\,F(\bx)e^{i\bk\cdot\bx}$ denotes the Fourier transform of the spatial smearing function $F(\bx)$, $\widetilde{\eta}_\bk(\tau)=\int^\tau_0\frac{\eta(s)}{\sqrt{2L^{n}\omega_\bk}}ds$, and the phase $\varphi_1=\sum_\bk\frac{2\lambda^2\widetilde{\eta}^2_\bk(\tau)|\widetilde{F}^\ast(\bk)|^2}{\omega_\bk\tau}$. Furthermore, the displacement operators 
$\hat{D}(\boldsymbol{\epsilon}_\bk)=\exp\big[\sum_\bk\big(\epsilon_\bk\hat{a}^\dagger_\bk-\epsilon^\ast_\bk\hat{a}_\bk\big)\big]$ 
and $\hat{D}\big(2\boldsymbol{\epsilon}_\bk\,e^{i\boldsymbol{\omega}_\bk\tau}\big)=\exp\big[\sum_\bk\big(2\epsilon_\bk\,e^{i\omega_\bk\tau}\hat{a}^\dagger_\bk-2\epsilon^\ast_\bk\,e^{-i\omega_\bk\tau}\hat{a}_\bk\big)\big]$ with the definition $\epsilon_\bk=\frac{\lambda\widetilde{\eta}_\bk(\tau)\widetilde{F}^\ast(\bk)}{\omega_\bk\tau}$.

Let us see what kind of operators we can obtain if considering the situation of $N$ pulse-sequence segment introduced above. By Hermitian conjugating and taking the operator to the $N$th power we can write $\hat{U}^\dagger_g=(\hat{u}^\dagger_g)^N$ as 
\begin{eqnarray}\label{Ug1}
\hat{U}^\dagger_g=e^{i\varphi_2}\hat{D}^\dagger(\boldsymbol{\epsilon}_\bk)\bigg[\hat{D}(2\boldsymbol{\epsilon}_\bk(1-e^{i\boldsymbol{\omega}_\bk\tau}))\exp\bigg[\sum_{\bk}\,2i\omega_{\bk}\tau\hat{a}^\dagger_\bk\hat{a}_\bk\bigg]\bigg]^N\hat{D}(\boldsymbol{\epsilon}_\bk).
\end{eqnarray}
Note that the phase $\varphi_2$ here incorporates $-N\varphi_1$ and an additional phase. However, the specific form of this additional phase
has no need to be given here, since the operator $\hat{U}_e$ carries exactly the opposite one. To further rewrite the operator \eqref{Ug1}, we 
will use the identities 
\begin{eqnarray}\label{ID}
\bigg[\hat{D}(x)e^{iy\hat{a}^\dagger\hat{a}}\bigg]^N=\bigg[\prod_{n=0}^{N-1}\hat{D}(xe^{iny})\bigg]e^{iNy\hat{a}^\dagger\hat{a}}
=e^{i\vartheta}\hat{D}\bigg(x\frac{1-e^{iNy}}{1-e^{iy}}\bigg)e^{iNy\hat{a}^\dagger\hat{a}},
\end{eqnarray}
to rewrite the powers term thereof. Here we also do not need to determine the phase $\vartheta$ because of the same reason as before. In doing so, $\hat{U}^\dagger_g$
in Eq. \eqref{Ug1} can be rewritten as 
\begin{eqnarray}\label{Ug2}
\hat{U}^\dagger_g=e^{i\varphi_3}\hat{D}^\dagger(\boldsymbol{\epsilon}_\bk)\hat{D}(\boldsymbol{\zeta}_\bk)\exp\bigg[\sum_{\bk}\,2iN\omega_{\bk}\tau\hat{a}^\dagger_\bk\hat{a}_\bk\bigg]\hat{D}(\boldsymbol{\epsilon}_\bk),
\end{eqnarray}
where $\hat{D}(\boldsymbol{\zeta}_\bk)=\hat{D}\big(2\boldsymbol{\epsilon}_\bk(1-e^{i\boldsymbol{\omega}_\bk\tau})
\frac{1-e^{2iN\boldsymbol{\omega}_\bk\tau}}{1-e^{2i\boldsymbol{\omega}_\bk\tau}}\big)$. Note that the phase $\varphi_3$ includes the phase $\varphi_2$ and the phase 
$\vartheta$ that is from identities \eqref{ID}. Similarly, we can rewrite $\hat{U}_e=(\hat{u}_e)^N$ as
\begin{eqnarray}\label{Ue2}
\hat{U}_e=e^{-i\varphi_3}\hat{D}^\dagger(\boldsymbol{\epsilon}_\bk)\hat{D}(\boldsymbol{\zeta}^\prime_\bk)\exp\bigg[\sum_{\bk}\,-2iN\omega_{\bk}\tau\hat{a}^\dagger_\bk\hat{a}_\bk\bigg]\hat{D}(\boldsymbol{\epsilon}_\bk),
\end{eqnarray}
with $\hat{D}(\boldsymbol{\zeta}^\prime_\bk)=\hat{D}\big(2\boldsymbol{\epsilon}_\bk(1-e^{-i\boldsymbol{\omega}_\bk\tau})
\frac{1-e^{-2iN\boldsymbol{\omega}_\bk\tau}}{1-e^{-2i\boldsymbol{\omega}_\bk\tau}}\big)$. 

Therefore, when $N$ pulse-sequence segment are applied  the corresponding evolution operator of the whole system is given by 
\begin{eqnarray}
\hat{U}(T)=|g\rangle\langle\,g|\hat{U}_g+|e\rangle\langle\,e|\hat{U}_e,
\end{eqnarray}  
where the total evolution time $T=2N\tau$ has been defined. Furthermore, using \eqref{Ug2} and \eqref{Ue2}, we can
calculate the relevant unitary operator $\hat{U}^\dagger_g\hat{U}_e$, which reads
\begin{eqnarray}\label{UgUe}
\hat{U}^\dagger_g\hat{U}_e=\hat{D}\bigg(-4\boldsymbol{\epsilon}_\bk\sin(N\boldsymbol{\omega}_\bk\tau)\tan\bigg(\frac{\boldsymbol{\omega}_\bk\tau}{2}\bigg)
e^{iN\boldsymbol{\omega}_\bk\tau}\bigg)=\hat{D}(\boldsymbol{\xi}_\bk(\tau, N)),
\end{eqnarray}
with $\boldsymbol{\xi}_\bk(\tau, N)$ being the arguments of the displacement operator. This unitary operator \eqref{UgUe} will be used later.

\section{The dynamic evolution of the auxiliary qubit}
At the beginning, the auxiliary qubit is prepared at its ground state $|g\rangle$, while the field is at an arbitrary state $\hat{\rho}_\text{field}$.
After the rotation, the auxiliary qubit is rotated to $|\psi(0)\rangle_\text{q}=\cos(\frac{\theta}{2})|g\rangle-i\sin(\frac{\theta}{2})|e\rangle$, while the field keeps its initial state. Then we apply a series of $2N~\pi$ pulses, all separated by the free evolution time $\tau$, and thereby modulate the effective Hamiltonian acting on the quantum field. 
Note that the total time of such evolution is $T=2N\tau$, and therefore the time-dependent state of the auxiliary qubit can be obtained by tracing over the degree of freedom of the quantum field:
\begin{eqnarray}\label{TES}
 \nonumber
\hat{\rho}_\text{q}(T)&=&\mathrm{Tr}_\text{field}\big[\hat{U}(T)|\psi(0)\rangle_\text{q}{_\text{q}}\langle\psi(0)|\otimes\hat{\rho}_\text{field}\hat{U}^\dagger(T)\big]
\\        \nonumber
&=&\mathrm{Tr}_\text{field}\bigg[\bigg(|g\rangle\langle\,g|\hat{U}_g+|e\rangle\langle\,e|\hat{U}_e\bigg)\bigg(\cos^2\frac{\theta}{2}|g\rangle\langle\,g|
+\sin^2\frac{\theta}{2}|e\rangle\langle\,e|+i\sin\frac{\theta}{2}\cos\frac{\theta}{2}|g\rangle\langle\,e|
-i\sin\frac{\theta}{2}\cos\frac{\theta}{2}|e\rangle\langle\,g|\bigg)
\\        \nonumber
&&\otimes\hat{\rho}_\text{field}
\bigg(|g\rangle\langle\,g|\hat{U}_g+|e\rangle\langle\,e|\hat{U}_e\bigg)^\dagger\bigg]
\\        \nonumber
&=&\mathrm{Tr}_\text{field}\bigg[\cos^2\frac{\theta}{2}|g\rangle\langle\,g|\otimes\hat{U}_g\hat{\rho}_\text{field}\hat{U}^\dagger_g+\sin^2\frac{\theta}{2}|e\rangle\langle\,e|\otimes\hat{U}_e\hat{\rho}_\text{field}\hat{U}^\dagger_e+i\sin\frac{\theta}{2}\cos\frac{\theta}{2}|g\rangle\langle\,e|
\otimes\hat{U}_g\hat{\rho}_\text{field}\hat{U}^\dagger_e
\\        \nonumber
&&
-i\sin\frac{\theta}{2}\cos\frac{\theta}{2}|e\rangle\langle\,g| \otimes\hat{U}_e\hat{\rho}_\text{field}\hat{U}^\dagger_g\bigg]
\\       \nonumber
&=&\cos^2\frac{\theta}{2}|g\rangle\langle\,g|+\sin^2\frac{\theta}{2}|e\rangle\langle\,e|+i\sin\frac{\theta}{2}\cos\frac{\theta}{2}\bigg[|g\rangle\langle\,e|
\mathrm{Tr}_\text{field}\big[\hat{U}_g\hat{\rho}_\text{field}\hat{U}^\dagger_e\big]-|e\rangle\langle\,g| 
\mathrm{Tr}_\text{field}\big[\hat{U}_e\hat{\rho}_\text{field}\hat{U}^\dagger_g\big]\bigg]
\\       \nonumber
&=&\cos^2\frac{\theta}{2}|g\rangle\langle\,g|+\sin^2\frac{\theta}{2}|e\rangle\langle\,e|+i\sin\frac{\theta}{2}\cos\frac{\theta}{2}\bigg[|g\rangle\langle\,e|
\mathrm{Tr}_\text{field}\big[\hat{\rho}_\text{field}\hat{U}^\dagger_e\hat{U}_g\big]-|e\rangle\langle\,g| 
\mathrm{Tr}_\text{field}\big[\hat{\rho}_\text{field}\hat{U}^\dagger_g\hat{U}_e\big]\bigg]
\\       \nonumber
&=&\cos^2\frac{\theta}{2}|g\rangle\langle\,g|+\sin^2\frac{\theta}{2}|e\rangle\langle\,e|+i\sin\frac{\theta}{2}\cos\frac{\theta}{2}\bigg[|g\rangle\langle\,e|
\mathrm{Tr}_\text{field}\big[\hat{D}^\dagger(\boldsymbol{\xi}_\bk(\tau, N))\hat{\rho}_\text{field}\big]-|e\rangle\langle\,g| 
\mathrm{Tr}_\text{field}\big[\hat{D}(\boldsymbol{\xi}_\bk(\tau, N))\hat{\rho}_\text{field}\big]\bigg]
\\       
&=&\frac{1}{2}\big[1-\cos(\theta)\sigma_z+\sin(\theta)(\Im[\chi(\boldsymbol{\xi}_\bk)]\sigma_x+\Re[\chi(\boldsymbol{\xi}_\bk)]\sigma_y)\big],
\end{eqnarray}
where $\chi(\boldsymbol{\xi}_\bk)=\mathrm{Tr}_\text{field}[\hat{\rho}_\text{field}\hat{D}(\boldsymbol{\xi}_\bk)]=\mathrm{Tr}_\text{field}\big[\hat{\rho}_\text{field}\exp\big\{\sum_\bk\big[\xi_\bk\hat{a}^\dagger_\bk-\xi^\ast_\bk\hat{a}_\bk\big]\big\}\big]$ is the characteristic function of the field corresponding to the state $\hat{\rho}_\text{field}$. The specific form of the argument of the displacement operator corresponding to mode $\bk$ reads 
\begin{eqnarray}
\nonumber
\xi_\bk(\tau, N)&=&-4\epsilon_\bk\sin(N\omega_\bk\tau)\tan\bigg(\frac{\omega_\bk\tau}{2}\bigg)
e^{iN\omega_\bk\tau}=-4\frac{\lambda\widetilde{\eta}_\bk(\tau)\widetilde{F}^\ast(\bk)}{\omega_\bk\tau}\sin(N\omega_\bk\tau)\tan\bigg(\frac{\omega_\bk\tau}{2}\bigg)
e^{iN\omega_\bk\tau}
\\      \nonumber
&=&-\frac{4\lambda\sin(N\omega_\bk\tau)\tan\big(\frac{\omega_\bk\tau}{2}\big)
e^{iN\omega_\bk\tau}}{\omega_\bk\tau\sqrt{2L^{n}\omega_\bk}}\int^\tau_0\eta(s)ds
\int\,d^n\bx\,F(\bx)e^{-i\bk\cdot\bx}
\\
&=&\lambda_\bk(\tau, N)\int^\tau_0\eta(s)ds\int\,d^n\bx\,F(\bx)e^{-i\bk\cdot\bx}.
\end{eqnarray}
Here, $\lambda_\bk(\tau, N)=-\frac{4\lambda\sin(N\omega_\bk\tau)\tan\big(\frac{\omega_\bk\tau}{2}\big)
e^{iN\omega_\bk\tau}}{\omega_\bk\tau\sqrt{2L^{n}\omega_\bk}}$ has been defined.

Seen from \eqref{TES} the expectation value of both the real and imaginary parts of the field displacement operator is mapped to the auxiliary qubit state. Through the proper measurements on the auxiliary qubit one thus can extract the full information about the symmetric characteristic function in principle. 
How to directly read out the symmetric characteristic function of quantum states of quantum field is shown in the main text.

%%%%%%%%%%%%%%%%%%%%%%%%%%%%%
\section{Characteristic function for different states of quantum scalar field}
Here we give the details about how to calculate the characteristic functions for a free scalar field in the Kubo-Martin-Schwinger (KMS) and squeezed states.

\subsection{Field in a finite-temperature KMS state}
Usually, the Gibbs thermal states are not well defined for QFT. Therefore, here we consider the field is prepared to be in a KMS state \cite{doi:10.1143/JPSJ.12.570, PhysRev.115.1342}, that generalizes Gibbs's notion of thermality to the quantum field case. 
More formally, for a KMS state $\hat{\rho}_\beta$ ($\beta$ is the inverse KMS temperature)
with respect to time translations generated by a Hamiltonian $\hat{H}$, the corresponding two-point correlator 
$\mathcal{W}_{\hat{\rho}_\beta}(\tau, \tau^\prime)=\mathrm{Tr}\{\hat{\rho}_\beta\hat{\phi}[t(\tau), \mathbf{x}(\tau)]\hat{\phi}[t(\tau^\prime), \mathbf{x}(\tau^\prime)]\}$ of the field satisfies the following two conditions: 1. $\mathcal{W}_{\hat{\rho}_\beta}(\tau, \tau^\prime)=\mathcal{W}_{\hat{\rho}_\beta}(\Delta\tau)$ with $\Delta\tau=\tau-\tau^\prime$ (stationarity), 
2. $\mathcal{W}_{\hat{\rho}_\beta}(\Delta\tau-i\beta)=\mathcal{W}_{\hat{\rho}_\beta}(-\Delta\tau)$ (antiperiodicity). The Wightman function from the KMS state of a 
free scalar field is well known\ \cite{PhysRevD.98.085007}
\begin{eqnarray}\label{Wightman-F}
\mathcal{W}_{\hat{\rho}_\beta}(t, \bx, t^\prime, \bx^\prime)=\mathcal{W}_{\hat{\rho}_\beta}(t-t^\prime, \bx-\bx^\prime)=\sum_\bk\frac{1}{e^{\beta\omega_\bk}-1}(e^{\beta\omega_\bk}e^{i\mathsf{k}\cdot(\mathsf{x}-\mathsf{x}^\prime)}
+e^{-i\mathsf{k}\cdot(\mathsf{x}-\mathsf{x}^\prime)}),
\end{eqnarray}
where 
\begin{eqnarray}
\mathrm{Tr}(\hat{\rho}_\beta\hat{a}_\bk\hat{a}^\dagger_{\bk^\prime})&=&\frac{e^{\beta\omega_\bk}}{e^{\beta\omega_\bk}-1}\delta^n(\bk-\bk^\prime),
\\
\mathrm{Tr}(\hat{\rho}_\beta\hat{a}^\dagger_{\bk}\hat{a}_{\bk^\prime})&=&\frac{1}{e^{\beta\omega_\bk}-1}\delta^n(\bk-\bk^\prime),
\\
\mathrm{Tr}(\hat{\rho}_\beta\hat{a}_{\bk}\hat{a}_{\bk^\prime})&=&0,
\\
\mathrm{Tr}(\hat{\rho}_\beta\hat{a}^\dagger_{\bk}\hat{a}^\dagger_{\bk^\prime})&=&0,
\end{eqnarray}
have been used to derive \eqref{Wightman-F}.

For the Gaussian state with zero one-point function, one can find \cite{deRamon:2020ljy}
\begin{eqnarray}
\langle\,e^{-i\hat{O}}\rangle=e^{-\frac{1}{2}\langle\hat{O}^2\rangle},
\end{eqnarray}
where $\hat{O}$ is an operator.
Therefore, using this identity we can calculate the characteristic function for thermal field state as
\begin{eqnarray}
\nonumber
\chi(\boldsymbol{\xi}_\bk)&=&\mathrm{Tr}_\text{field}[\hat{\rho}_\beta\hat{D}(\boldsymbol{\xi}_\bk)]=\mathrm{Tr}_\text{field}\bigg[\hat{\rho}_\beta\exp\bigg\{\sum_\bk\big[\xi_\bk\hat{a}^\dagger_\bk-\xi^\ast_\bk\hat{a}_\bk\big]\bigg\}\bigg]
\\         \nonumber  
&=&\exp\bigg[\frac{1}{2}\mathrm{Tr}_\text{field}\bigg\{\hat{\rho}_\beta\sum_\bk\sum_{\bk^\prime}[\xi_\bk\hat{a}^\dagger_\bk-\xi^\ast_\bk\hat{a}_\bk\big]
\big[\xi_{\bk^\prime}\hat{a}^\dagger_{\bk^\prime}-\xi^\ast_{\bk^\prime}\hat{a}_{\bk^\prime}\big]\bigg\}\bigg]
\\         \nonumber  
&=&\exp\bigg[-\frac{1}{2}\sum_\bk|\xi_\bk|^2\frac{e^{\beta\omega_\bk}+1}{e^{\beta\omega_\bk}-1}\bigg].
\end{eqnarray}
The covariance matrix of thermal state of mode $\bk$ is $V_\bk=(2n_\bk+1)\mathbb{I}$ with $n_\bk=1/(e^{\beta\omega_\bk}-1)$. Using this, the above characteristic 
function can be rewritten as
 \begin{eqnarray}\label{chi}
\chi(\boldsymbol{\xi}_\bk)=\exp\bigg[-\frac{1}{2}\sum_\bk\xi_\bk^T(\Omega\,V_\bk\Omega^T)\xi_\bk\bigg]
=\exp\bigg[-\frac{1}{2}\boldsymbol{\xi}_\bk^T(\boldsymbol{\Omega}\boldsymbol{V}_\bk\boldsymbol{\Omega}^T)\boldsymbol{\xi}_\bk\bigg],
\end{eqnarray}
where $\boldsymbol{\xi}_\bk=(\xi_{k_1}, \dots, \xi_{k_n})^T$, $\boldsymbol{\Omega}=\bigoplus_{i=1}^n\Omega_i$ (here $\Omega=\begin{pmatrix}
0   &    1       \\
-1  &    0
\end{pmatrix}
$ known as the symplectic matrix \cite{ferraro2005gaussian, RevModPhys.84.621}), and $\boldsymbol{V}_\bk=\bigoplus_{i=1}^nV_{k_i}$
have been used. Note that each $\xi_{k_i}$ actually contains both real and imaginary parties written as $\xi_{k_i}=(\Re(\xi_{k_i}), \Im(\xi_{k_i}) )^T$.

%%%%%%%%%%%%%%%%%%%%%%%%%%%%%%%%%%%%%%%%%%%%%%%%%%%%
\subsection{Field in a squeezed state}
We consider a squeezed state for the scalar field without mixed squeezing between different modes. More formally, it is of the form,
\begin{eqnarray}
|\boldsymbol{\zeta}\rangle=\hat{S}_{\boldsymbol{\zeta}}|0\rangle=\exp\bigg[\frac{1}{2}\sum_\bk\big(\zeta^\ast_\bk\hat{a}^2_\bk-\zeta_\bk\hat{a}^{\dagger2}_\bk\big)\bigg]|0\rangle.
\end{eqnarray}
Here $\zeta_\bk=r_\bk\,e^{i\theta_\bk}$ written in the polar form is the squeezing parameters for different modes of the field.
The squeezing transformation of the annihilation and creation operators, $\hat{a}_\bk$ and $\hat{a}^\dagger_\bk$, is given by 
\begin{eqnarray}\label{ST}
\hat{S}^\dagger_{\boldsymbol{\zeta}}\hat{a}_\bk\hat{S}_{\boldsymbol{\zeta}}&=&\cosh[\zeta_\bk]\hat{a}_\bk-e^{i\theta_\bk}\sinh[\zeta_\bk]\hat{a}^\dagger_\bk,
\\
\hat{S}^\dagger_{\boldsymbol{\zeta}}\hat{a}^\dagger_\bk\hat{S}_{\boldsymbol{\zeta}}&=&\cosh[\zeta_\bk]\hat{a}^\dagger_\bk-e^{-i\theta_\bk}\sinh[\zeta_\bk]\hat{a}_\bk,
\end{eqnarray}
Through this squeezing transformation one can find 
\begin{eqnarray}
\mathrm{Tr}(|\boldsymbol{\zeta}\rangle\langle\boldsymbol{\zeta}|\hat{a}_\bk\hat{a}^\dagger_{\bk^\prime})&=&\cosh^2[\zeta_\bk]\delta^n(\bk-\bk^\prime),
\\
\mathrm{Tr}(|\boldsymbol{\zeta}\rangle\langle\boldsymbol{\zeta}|\hat{a}^\dagger_{\bk}\hat{a}_{\bk^\prime})&=&\sinh^2[\zeta_\bk]\delta^n(\bk-\bk^\prime),
\\
\mathrm{Tr}(|\boldsymbol{\zeta}\rangle\langle\boldsymbol{\zeta}|\hat{a}_{\bk}\hat{a}_{\bk^\prime})&=&-\cosh[\zeta_\bk]\sinh[\zeta_\bk]e^{i\theta_\bk}\delta^n(\bk-\bk^\prime),
\\
\mathrm{Tr}(|\boldsymbol{\zeta}\rangle\langle\boldsymbol{\zeta}|\hat{a}^\dagger_{\bk}\hat{a}^\dagger_{\bk^\prime})&=&
-\cosh[\zeta_\bk]\sinh[\zeta_\bk]e^{-i\theta_\bk}\delta^n(\bk-\bk^\prime).
\end{eqnarray}

\iffalse
The Wightman function from this squeezed state of 
a scalar field is given by \cite{PhysRevD.98.085007}
\begin{eqnarray}
\nonumber 
\mathcal{W}_{\hat{\rho}_{\boldsymbol{\zeta}}}(t, \bx, t^\prime, \bx^\prime)&=&\sum_\bk\big[\cosh^2[\zeta_\bk]e^{i\mathsf{k}\cdot(\mathsf{x}-\mathsf{x}^\prime)}
+\sinh^2[\zeta_\bk]e^{-i\mathsf{k}\cdot(\mathsf{x}-\mathsf{x}^\prime)}-e^{i\theta_\bk}\sinh[\zeta_\bk]\cosh[\zeta_\bk]e^{-i\omega_\bk(t+t^\prime)}e^{i\bk\cdot(\bx+\bx^\prime)}
\\       \nonumber
&&-e^{-i\theta_\bk}\sinh[\zeta_\bk]\cosh[\zeta_\bk]e^{i\omega_\bk(t+t^\prime)}e^{-i\bk\cdot(\bx+\bx^\prime)}\big].
\end{eqnarray}
Note that, unlike the KMS state case, the Wightman function for a squeezed state is not invariant with respective to spacetime translations as a result of the appearance 
the factors $t+t^\prime$ and $\bx+\bx^\prime$. 
\fi

Similar to the analysis of the KMS state case, we can also find the characteristic function for a free scalar field in the squeezed state,
\begin{eqnarray}
\nonumber
\chi(\boldsymbol{\xi}_\bk)&=&\mathrm{Tr}_\text{field}[|\boldsymbol{\zeta}\rangle\langle\boldsymbol{\zeta}|\hat{D}(\boldsymbol{\xi}_\bk)]=\mathrm{Tr}_\text{field}\bigg[|\boldsymbol{\zeta}\rangle\langle\boldsymbol{\zeta}|\exp\bigg\{\sum_\bk\big[\xi_\bk\hat{a}^\dagger_\bk-\xi^\ast_\bk\hat{a}_\bk\big]\bigg\}\bigg]
\\         \nonumber  
&=&\exp\bigg[\frac{1}{2}\mathrm{Tr}_\text{field}\bigg\{|\boldsymbol{\zeta}\rangle\langle\boldsymbol{\zeta}|\sum_\bk\sum_{\bk^\prime}[\xi_\bk\hat{a}^\dagger_\bk-\xi^\ast_\bk\hat{a}_\bk\big]
\big[\xi_{\bk^\prime}\hat{a}^\dagger_{\bk^\prime}-\xi^\ast_{\bk^\prime}\hat{a}_{\bk^\prime}\big]\bigg\}\bigg]
\\        
&=&\exp\bigg[-\frac{1}{2}\sum_\bk\big(\cosh[2\zeta_\bk]|\xi_\bk|^2+\sinh[2\zeta_\bk]\Re\big[e^{i\theta_\bk}\xi^{\ast2}_\bk\big]\big)\bigg].
\end{eqnarray}
The covariance matrix of the squeezed state of mode $\bk$ reads
\begin{eqnarray}
V_\bk=
\begin{pmatrix}
\cosh[2\zeta_\bk]-\cos\theta_\bk\sinh[2\zeta_\bk]   &    -\sin\theta_\bk\sinh[2\zeta_\bk]       \\
-\sin\theta_\bk\sinh[2\zeta_\bk]  &    \cosh[2\zeta_\bk]+\cos\theta_\bk\sinh[2\zeta_\bk] 
\end{pmatrix}.
\end{eqnarray}
Then the above characteristic function  can be rewritten as
 \begin{eqnarray}\label{chi}
\chi(\boldsymbol{\xi}_\bk)=\exp\bigg[-\frac{1}{2}\sum_\bk\xi_\bk^T(\Omega\,V_\bk\Omega^T)\xi_\bk\bigg]
=\exp\bigg[-\frac{1}{2}\boldsymbol{\xi}_\bk^T(\boldsymbol{\Omega}\boldsymbol{V}_\bk\boldsymbol{\Omega}^T)\boldsymbol{\xi}_\bk\bigg].
\end{eqnarray}

\end{document}